\documentclass[prb,twocolumn,nopacs,floatfix,superscriptaddress,amsmath]{revtex4-2}        

\usepackage{amssymb}
\usepackage{amsmath}       %prl-be ez kell
\usepackage{amstext} 
\usepackage{graphicx}
\usepackage{xcolor}
\usepackage{enumerate}
\usepackage[normalem]{ulem}
\usepackage[colorlinks,bookmarks=false,citecolor=blue,linkcolor=blue,urlcolor=blue]{hyperref}
\usepackage{xspace}
\usepackage[capitalize]{cleveref}

\newcommand{\rucl}{\mbox{$\alpha$-RuCl$_3$}\xspace}
\newcommand{\cstar}{\ensuremath{{c^{\ast}}}}
\newcommand{\eps}{\ensuremath{\epsilon_\cstar}}
\renewcommand{\eps}{\ensuremath{\epsilon}}
\newcommand{\compressibility}{\kappa_{\cstar}}

\newcommand{\MEC}[1]{\ensuremath{\widetilde{#1}}}

\begin{document}

\linepenalty=10
\interlinepenalty=100
\clubpenalty=10000
\widowpenalty=10000
\brokenpenalty=1000
\tolerance=900
\hyphenpenalty=5000

%Giving a title
\title{\boldmath Investigation of the magnetoelastic coupling anisotropy in the Kitaev material \rucl \unboldmath}

%Giving the authors and affiliation
\author{Vilmos Kocsis}
\affiliation{Institut f\"ur Festk\"orperforschung, Leibniz IFW-Dresden, 01069 Dresden, Germany}

\author{David A. S. Kaib}
\affiliation{Institut f\"ur Theoretische Physik, Goethe-Universit\"at Frankfurt, Max-von-Laue-Strasse 1, 60438 Frankfurt am Main, Germany}

\author{Kira Riedl}
\affiliation{Institut f\"ur Theoretische Physik, Goethe-Universit\"at Frankfurt, Max-von-Laue-Strasse 1, 60438 Frankfurt am Main, Germany}

\author{Sebastian Gass}
\affiliation{Institut f\"ur Festk\"orperforschung, Leibniz IFW-Dresden, 01069 Dresden, Germany}

\author{Paula Lampen-Kelley}
\affiliation{Department of Materials Science and Engineering, University of Tennessee, Knoxville, TN 37996, U.S.A.}
\affiliation{Materials Science and Technology Division, Oak Ridge National Laboratory, Oak Ridge, TN 37831, U.S.A.}

\author{David G. Mandrus}
\affiliation{Department of Materials Science and Engineering, University of Tennessee, Knoxville, TN 37996, U.S.A.}
\affiliation{Materials Science and Technology Division, Oak Ridge National Laboratory, Oak Ridge, TN 37831, U.S.A.}

\author{Stephen E. Nagler}
\affiliation{Neutron Scattering Division, Oak Ridge National Laboratory, Oak Ridge, TN 37831, U.S.A.}

\author{Nicol\'as P\'erez}
\affiliation{Institut f\"ur Metallische Werkstoffe, Leibniz IFW-Dresden, 01069 Dresden, Germany}

\author{Kornelius Nielsch}
\affiliation{Institut f\"ur Metallische Werkstoffe, Leibniz IFW-Dresden, 01069 Dresden, Germany}

\author{Bernd B\"uchner}
\affiliation{Institut f\"ur Festk\"orperforschung, Leibniz IFW-Dresden, 01069 Dresden, Germany}
\affiliation{Institut für Festk\"orper- und Materialphysik and W\"urzburg-Dresden Cluster of Excellence ct.qmat, Technische Universit\"at Dresden, 01062 Dresden, Germany}

\author{Anja U. B. Wolter}
\affiliation{Institut f\"ur Festk\"orperforschung, Leibniz IFW-Dresden, 01069 Dresden, Germany}

\author{Roser Valent\'i}
\affiliation{Institut f\"ur Theoretische Physik, Goethe-Universit\"at Frankfurt, Max-von-Laue-Strasse 1, 60438 Frankfurt am Main, Germany}

% PACS numbers

\begin{abstract}
The Kitaev material \rucl is among the most prominent candidates to host a quantum spin-liquid state endowed with fractionalized excitations.
Recent experimental and theoretical investigations have separately revealed the importance of both the magnetoelastic coupling and the magnetic anisotropy, in dependence of the applied magnetic field direction.
In this combined theoretical and experimental research, we investigate the anisotropic magnetic and magnetoelastic properties for magnetic fields applied along the main crystallographic axes as well as for fields canted out of the honeycomb plane.
We found that the magnetostriction anisotropy is unusually large compared to the anisotropy of the magnetization, which is related to the strong magnetoelastic $\MEC{\Gamma'}$-type coupling in our \textit{ab-initio} derived model.
We observed large, non-symmetric magnetic anisotropy for magnetic fields canted out of the honeycomb $ab$-plane in opposite directions, namely towards the $+c^*$ or $-c^*$ axes, respectively.
The observed directional anisotropy is explained by considering the relative orientation of the magnetic field with respect to the co-aligned RuCl$_6$ octahedra.
Magnetostriction measurements in canted fields support this non-symmetric magnetic anisotropy, however these experiments are affected by magnetic torque effects.
Comparison of theoretical predictions with experimental findings allow us to recognize the significant contribution of torque effects in experimental setups where \rucl is placed in canted magnetic fields.
\end{abstract}

\maketitle

%%%%%%%%%%%%%%%%%%%%%%%%%%%%%%%%%%%%%%%%%%%%%%%%%%%%%%%%%%%%%%%%%%%%%%%%%%%%%
%
%
%%%%%%%%%%%%%%%%%%%%%%%%%%%%%%%%%%%%%%%%%%%%%%%%%%%%%%%%%%%%%%%%%%%%%%%%%%%%%

	\section{Introduction}

Materials hosting quantum spin-liquid states have attracted much interest recently~\cite{Sachdev2008NPhys,Balents2010,Norman2016RMP,rau2016anurev,Winter2017JPCM,Takagi2019NRP,Broholm2020Sci}, as in these systems the quantum information may be protected from decoherence, and they can be applied in quantum computing technology~\cite{Nayak2008RMP}.
A prime example for a theoretical model to host a quantum spin-liquid state is provided by the exactly solvable Kitaev model on the honeycomb lattice~\cite{Kitaev2006}, which contains frustrated, bond-dependent magnetic interactions that lead to fractionalized quasiparticles; gauge fluxes and Majorana fermions.
The investigation and experimental verification of quantum spin-liquid states presents, however, an ongoing challenge, that has brought \rucl to the forefront of research as a prime candidate for Kitaev physics.

While the honeycomb-layered \rucl orders antiferromagnetically at low temperatures~\cite{Johnson2015PRB,Sears2015PRB}, the possibility of residual physics of fractionalization~\cite{nasu2016fermionic,Do2017NatPhys,Jansa2018NatPhys,Motome2020JPSJ,Li2020PRR} or even a field-induced Kitaev spin-liquid state~\cite{yadav2016KitaevExchangeFieldinduced,Baek2017PRL,Banerjee2018QM} have been intensively discussed.
So far, numerous experimental methods have been used for the investigation of \rucl, including neutron and Raman scattering~\cite{Sandilands2015PRL,Do2017NatPhys,Banerjee2018QM,Balz2019PRB,sahasrabudhe2020HighfieldQuantumDisordered,Wulferding2020NatComm,Wang2020QM}, specific heat~\cite{Wolter2017PRB}, Grüneisen parameter~\cite{bachus2020Thermodynamic}, microwave and terahertz absorption~\cite{Baek2017PRL,Wang2017PRL,Wellm2018PRB}, as well as thermal transport measurements~\cite{KasaharaPRL2018,Hentrich2019PRB,HentrichPRL2019,czajka2021OscillationsThermalConductivitya}, and notably some reporting a half-integer-quantized thermal Hall conductivity~\cite{Kasahara2018Nat,yokoi2020half,Yamashita2020PRB,bruin2021RobustnessThermalHall}. 

Various Raman scattering studies have reported pronounced Fano line shapes~\cite{Sandilands2015PRL,Mai2019PRB,sahasrabudhe2020HighfieldQuantumDisordered,Wulferding2020NatComm}, which evidence a significant magnetoelastic coupling between the phonon modes and the magnetic continuum.
Indeed, more recent thermal expansion and magnetostriction measurements have probed direct consequences of such coupling in \rucl~\cite{He2018JPCM,Gass2020PRB,Schonemann2020PRB}.
Strong magnetostrictive effects are plausible, considering those two aspects.
Firstly, magnetoelastic coupling is expected to be especially sensitive in Kitaev materials due to the strongly geometry-dependent exchange mechanisms~\cite{jackeli2009MottInsulatorsStrong,rau2016anurev,Winter2016PRB}.   
Secondly, the weak van-der-Waals force between the honeycomb layers leads to large changes in the lattice parameters when mechanical stress is applied.
We note that magnetoelastic coupling is necessarily in play when measuring a hypothetical spinful chiral edge current (that would be present in the Kitaev spin-liquid) \cite{vinkler-aviv2018ApproximatelyQuantizedThermal,ye2018QuantizationThermalHall}. 
Furthermore, magnetoelastic coupling could also lend the phonons themselves a bulk transverse (Hall) current, see e.g. Ref.~\onlinecite{ye2021phonon}.

Further understanding of the intrinsic anisotropy in \rucl\ could be gained by field-angular dependent measurements. 
So far, significant magnetic torque effects have been found and investigated for various directions of the magnetic ($H$) field \cite{Leahy2017PRL,modic2018ResonantTorsionMagnetometry,modic2018ChiralSpinOrder,Riedl2019PRL}. 
Additionally, specific heat and thermal conductivity measurements have been performed in magnetic fields applied in various in-plane and out-of-plane angles and revealed anisotropic thermodynamic and transport properties \cite{Kasahara2018Nat,yokoi2020half,Yamashita2020PRB,czajka2021OscillationsThermalConductivitya,bruin2021RobustnessThermalHall,tanaka2020thermodynamic}.
Therefore, combined investigations of the magnetoelastic coupling and the magnetic anisotropy, using canted fields (i.e. fields tilted out of the honeycomb plane), can help to unveil the complex behavior of \rucl.

In this combined experimental and theoretical study, we focus on the angular, temperature, and magnetic field dependence of the magnetic and magnetoelastic properties of \rucl. 
Depending on the in-plane field angle, we resolve a phase transition between different antiferromagnetic orders, in accord with recent previous studies.
In the presence of canted magnetic fields, we observe an anomalous increase in the magnetostriction at high fields related to the magnetic torque effects.
The combination of magnetic measurements reveals a significant, non-symmetric anisotropy for magnetic fields canted out of the hexagonal $ab$-plane in opposite directions, upwards or downwards, namely towards the $+c^*$ or $-c^*$ axes.
This angular-anisotropy is related to the co-aligned, corner-sharing RuCl$_6$ octahedra within the hexagonal planes.
The experimentally observed magnetic and magnetoelastic anisotropy is the largest when the $H$ field is rotated within the $ac^*$ plane, and smallest when rotated within the $bc^*$ plane.
To model the magnetostriction and the effect of magnetic torque, we employ \textit{ab-initio} derived magnetoelastic couplings \cite{kaib2021magnetoelastic}, allowing us to separate the different contributions of the magnetoelastic interactions.
The theoretical model provides a good qualitative description of the experimental observations, namely predicting non-symmetric angular-anisotropy for the $ac^*$ plane while excluding it for the $bc^*$ plane.
However, our experiments and the theoretical model also point out the significant role of magnetic torque in those experiments, where the sample can freely move or deform, such as in magnetostriction and thermal transport measurements.
In case of magnetostriction measurement, the movement or deformation of the sample is on the sub-$\mu$m scale, while in case of thermal transport measurements the deformation can be significantly higher.

%%%%%%%%%%%%%%%%%%%%%%%%%%%%%%%%%%%%%%%%%%%%%%%%%%%%%%%%%%%%%%%%%%%%%%%%%%%%%
%
%
%%%%%%%%%%%%%%%%%%%%%%%%%%%%%%%%%%%%%%%%%%%%%%%%%%%%%%%%%%%%%%%%%%%%%%%%%%%%%
	\section{Experimental and Theoretical Methods}
	\subsection{Experimental details}
	Single crystals of \rucl were grown using the chemical vapor transport method~\cite{Banerjee2017Sci}.
	The orientations of the monoclinic $a$ and $b$ axes with respect to the honeycomb plane were determined by angular dependent magnetization measurements with $\mathbf{H}\in{ab}$ fields.
	The angular dependent magnetization measurements were carried out in a SQUID magnetometer (MPMS-XL, Quantum Design).
	The field dependent magnetization measurements up to $\mu_0H$=14\,T were measured in a vibrating sample magnetometer (VSM, PPMS, Quantum Design).
	The precise 45\,deg canting orientation of the crystals was ensured by a pair of appropriately cut quartz pads, between which the sample was fixed with varnish.
	The $ab$-plane orientation of the crystals was aligned under a microscope with $\pm$1-2\,deg angular precision.

\begin{figure}
    \includegraphics[width=8.5truecm]
    {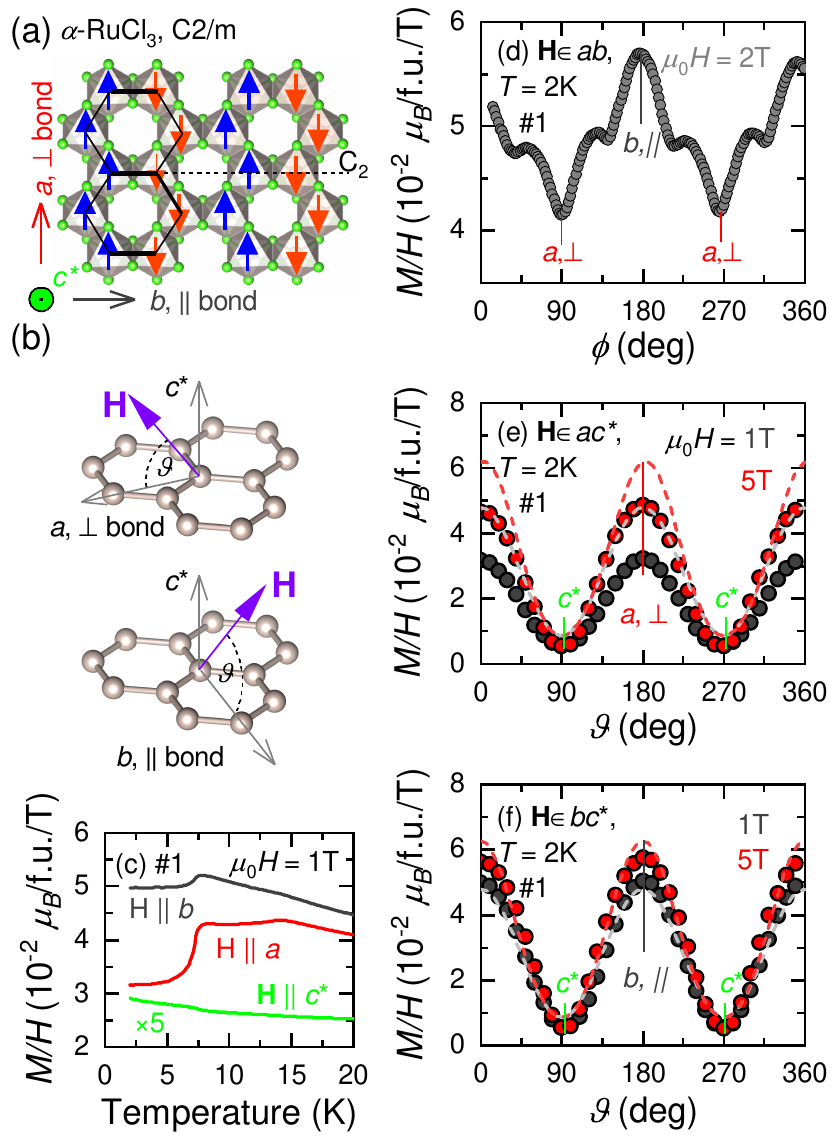}
    \caption{(Color online)
    (a)~Single honeycomb layer of \rucl and the crystallographic axes ($a$, $b$, and $c^*$). We highlight two directions within the $ab$ plane; the $a$ axis is perpendicular ($\perp{bond}$) and the $b$ axis is parallel ($\parallel{bond}$) to one of the Ru-Ru bonds, respectively. Red and blue arrows at the honeycomb sites indicate the zigzag domain with ordering wave vector $\mathbf Q \parallel b$. Dashed line indicates a C$_2$ rotation symmetry around an axis parallel to the $b$ axis.
    (b)~During the magnetization and magnetostriction measurements, the magnetic field ($\mathbf{H}$) was canted out of the $ab$ plane by an angle of $\vartheta$, while the planar projection of the applied field was either along the $a$ or $b$ axis.
    (c)~Temperature dependence of the magnetization for fields along the main crystallographic axes in the field cooling runs ($\mu_0H$=1\,T). Note, that the data for $\mathbf{H}\parallel{c^*}$ is multiplied by a factor of 5 for better visibility.
    (d-e)~Angular dependence of the magnetization at $T$=2\,K for fields rotated within the $ab$, $ac^*$, and $bc^*$ planes, respectively. Measurement data is plotted with symbols (full circles) for $\mu_0H$=1\,T and 5\,T. In case of $\mathbf{H}\in{ab}$, the measurements are plotted for $\mu_0H$=2\,T. Dashed curves in panels (e,f) correspond to the theoretical calculations.
    }
    \label{RuCl3-1}
    \end{figure}

The magnetostriction was measured using a custom-built dilatometer based on the capacitance measurement technique (AH2700A, Andeen-Hagerling)~\cite{Pott1983}.
Due to the dimensions of the available single crystals, the length change $\Delta{L}$ of the sample was measured along the $c^*$ axis ($\Delta{L}_{c^*}\parallel{c}^*$, see \cref{RuCl3-1}(a)), while the $H$ field could be applied in arbitrary directions via the rotation of the capacitance cell body or the sample.
In this measurement technique, the sample is held in place in the dilatometer by a small uniaxial pressure applied on the sample during the mounting.
Therefore if sufficient torque is applied, the sample may slightly rotate or deform within the dilatometer, which is measured as an apparent length change.
This issue will be discussed in details in Sec.~\ref{sec:MandME} as well as we give an estimate to its magnitude.
During the magnetostriction measurement the magnetic field was swept between $\pm$14\,T with 0.01\,T/min or 0.03\,T/min rates at constant temperatures.
The linear magnetostriction coefficient along the $c^*$ axis ($\lambda_{c^*}$) was calculated as the $H$-field derivative of the relative length change:
\begin{equation}
\lambda_{c^*}=\frac{\partial}{\partial(\mu_0H)}\frac{\Delta{L}_{c^*}(T,\mu_0H)}{L_{c^*}(300\,K,0\,T)}. \label{eq:magnetostrictionexp}
\end{equation}
The measurements were performed on two different pieces of \rucl crystals from the same batch (samples $\#$1 and $\#2$) with thicknesses of $\sim$800\,$\mu$m.

\subsection{Theoretical details \label{sec:theory}}
We compare our measurements to numerical results on the extended Kitaev models.
In such models, the bonds are labeled as X, Y or Z depending on their orientation. 
For a nearest-neighbor Z-bond (parallel to the $b$ axis, see \cref{RuCl3-1}(a)) with local $C_{2h}$ symmetry, the symmetry-allowed magnetic exchange between the $J_{\text{eff}}=\frac12$ pseudospins, labeled as $\mathbf S_i$ and $\mathbf S_j$ is \cite{rau2014generic}
\begin{align}
    H_{\mathrm Z} = & K S_{i}^{z}  S_{j}^{z} + J \mathbf{S}_{i} \cdot \mathbf{S}_{j} +\Gamma\left(S_{i}^{x} S_{j}^{y}+S_{i}^{y} S_{j}^{x}\right) \notag \\
   & +\Gamma^{\prime}\left(S_{i}^{x} S_{j}^{z}+S_{i}^{z} S_{j}^{x}+S_{i}^{y} S_{j}^{z}+S_{i}^{z} S_{j}^{y}\right), \label{eq:Hamil_mag}
\end{align}
where $K$ and $J$ correspond to the Kitaev and Heisenberg exchanges, respectively, while $\Gamma, \Gamma'$ are symmetric off-diagonal exchanges.
The X and Y bond exchanges can be constructed via the cyclic permutation of $(x,y,z)$ in \cref{eq:Hamil_mag}. The magnetic Hamiltonian is then given as the sum of these exchange terms (including possible longer-range terms) and the Zeeman term $H_\text{Zee}=-\mu_B\mu_0 \sum_i \mathbf H \cdot \mathbb G \cdot \mathbf S_i$, where $\mathbb G$ is the gyromagnetic tensor.
To solve it, we employ exact diagonalization (ED) on a hexagon-shaped 24-site cluster. 
As a magnetic model, we discuss the \textit{ab-initio} guided minimal model of Ref.~\onlinecite{winte17}, which has been shown to reproduce many experimental observations in \rucl~\cite{winte17,Wolter2017PRB,winte18,cookmeyer2018SpinwaveAnalysisLowtemperature,Riedl2019PRL,sahasrabudhe2020HighfieldQuantumDisordered,bachus2020Thermodynamic,bachus2021angle}.
Here the exchange parameters are 
\begin{equation}
(K,\,J,\,\Gamma,\,\Gamma',\,J_3)=(-5,\,-0.5,\,2.5,\,0,\,0.5)\text{\,meV},
\end{equation}
where $J_3$ denotes an additional third-nearest-neighbor Heisenberg exchange, and the components of $\mathbb G$ are $g_{ab}=2.3$ and $g_\cstar =1.3$, for the in-plane and out-of-plane elements, respectively.
Note that this model is $C_3$-simplified, \textit{i.e.} the coupling magnitudes are equal on X, Y and Z bonds.
The $C2/m$ structure of \rucl \cite{Johnson2015PRB} does however slightly break $C_3$ symmetry, a property that manifests in the \textit{in-plane} angle-dependent measurements discussed below and therefore it is not described in the present model by construction.

To model the spin-lattice coupling, we employ the \textit{ab-initio}-derived linear magnetoelastic couplings of Ref.~\onlinecite{kaib2021magnetoelastic} for \rucl, defined as $\MEC{\mathcal J}=\left(\frac{\partial \mathcal J}{\partial \eps}\right)|_{\eps=0}$, where $\eps = \Delta {L_\cstar}/L_{\cstar}$ and $\mathcal{J} \in \{{K},{J},\dots,g_{ab},g_\cstar\}$.
The strongest magnetoelastic exchange couplings are then 
\begin{equation}
(\MEC{K},\MEC{J},\MEC{\Gamma},\MEC{\Gamma'})=(40.5,\,1.3,\,7.5,\,-11.5)\text{\,meV} \label{eq:MECmodel}
\end{equation}
and the magnetoelastic $g$ couplings $(\MEC{g_{ab}},\MEC{g_{\cstar}})=(-1.6,\,3.85)$.
Note that the predicted large magnetoelastic $\MEC\Gamma'$ coupling in this model is a somewhat unexpected property, as the \textit{magnetic} $\Gamma'$ coupling is generally found to be subdominant or negligible in magnetic models of \rucl (see, e.g., Ref.~\onlinecite{laurell2020DynamicalThermalMagnetic}).
Nevertheless we find this large $\MEC \Gamma'$ to be essential to reproduce the strong anisotropy found in our magnetostriction measurements, as discussed below. 
In our calculations we also include the weaker longer-range magnetoelastic couplings of Ref.~\onlinecite{kaib2021magnetoelastic}, which however do not qualitatively change the results. For the magnetostriction, we then employ the approximation \cite{kaib2021magnetoelastic}
\begin{equation}
    \lambda_{c^*} \approx \frac{\compressibility}{V} \sum_{\mathcal J\in\{K,J,\dots\}} \MEC{\mathcal J} \, \left(\frac{\partial M}{\partial \mathcal J} \right)_{\eps=0}, \label{eq:lambda_theory}
\end{equation}
where the sum goes through all strain-dependent interactions and $g$ values. 
The parameter $\compressibility \equiv - (\partial  \eps/\partial p_\cstar)$ is the (unknown) linear compressibility along $\cstar$ against uniaxial pressure $p_\cstar$.
The field-dependence of $\lambda_\cstar$ enters through the field-dependencies of the magnetization susceptibilities $\left(\frac{\partial M}{\partial \mathcal J} \right)_{\eps=0}$, which we compute using  ED in the magnetic model described above. 

%%%%%%%%%%%%%%%%%%%%%%%%%%%%%%%%%%%%%%%%%%%%%%%%%%%%%%%%%%%%%%%%%%%%%%%%%%%%%
%
%
%%%%%%%%%%%%%%%%%%%%%%%%%%%%%%%%%%%%%%%%%%%%%%%%%%%%%%%%%%%%%%%%%%%%%%%%%%%%%
\section{Magnetic and elastic properties in canted magnetic fields} \label{sec:MandME}
Each layer of \rucl consists of edge-sharing RuCl$_6$ octahedra that form a honeycomb network, as shown in Fig.~\ref{RuCl3-1}(a).
For the crystal structure, both the rhomboedral $R\bar 3$ \cite{park2016emergence,glamazda2017RelationKitaevMagnetism,janssen2020MagnonDispersionDynamic} and the monoclinic $C2/m$ \cite{Johnson2015PRB,cao2016low} structures are presently discussed in the literature.
We employ the axis convention of the $C2/m$ structure, where the honeycomb plane is spanned by the crystallographic $a$ and $b$ axes, while $\cstar$ is perpendicular to it, see \cref{RuCl3-1}(a,b). Note that the $b$ axis is parallel to one of the honeycomb bonds, while the $a$ axis is perpendicular to the same bond.

The antiferromagnetic \lq\lq{zigzag}\rq\rq\  long-range order \cite{Johnson2015PRB} (Fig.~\ref{RuCl3-1}(a)) develops at $T_{\rm N}$=7.1\,K, as shown by the magnetization data (Fig.~\ref{RuCl3-1}(c)) in moderate $\mu_0H$=1\,T fields applied along the main crystallographic axes.
The magnetization curves for $\mathbf{H}\parallel{a}$ and $\mathbf{H}\parallel{b}$ show a sudden decrease at $T_{\rm N}$, however the weaker temperature dependence for $\mathbf{H}\parallel{b}$ suggests that the ordered moments are perpendicular to the $b$ axis.
The particular zigzag domain structure associated to such ordering  \cite{chaloupka2016MagneticAnisotropyKitaev}, where the ordered moments lie in the $a\cstar$-plane, is  illustrated in Fig.~\ref{RuCl3-1}(a).
Note, that the minor transition apparent for $\mathbf{H}\parallel{a}$ at $T$=14\,K is indicative of the so-called ABC/ABAB-stacking faults~\cite{Sears2015PRB,Banerjee2016NatMat}. 

For fields applied perpendicular to the honeycomb plane ($\mathbf H \parallel \cstar$) a much smaller susceptibility is found, highlighting the strong easy-plane anisotropy in \rucl.
This is further resolved in \cref{RuCl3-1}(e) and \ref{RuCl3-1}(f), where the field is rotated within the $a\cstar$-plane or $b\cstar$-plane, respectively (cf.~\cref{RuCl3-1}(b)), in the presence of constant field strength and temperature $T$=2\,K.
Corresponding theoretical $T$=0\,K results within the magnetic minimal model (see \cref{sec:theory}) agree well with the measurement, see dashed lines in \cref{RuCl3-1}(e,f).
In case of magnetic properties, the easy-plane anisotropy is primarily facilitated by the strong $\Gamma$-term and the anisotropic $g$-tensor~\cite{janssen2017MagnetizationProcessesZigzag,Riedl2019PRL}. 
Note, that theoretical curves in Figs.~\ref{RuCl3-1}(e) and \ref{RuCl3-1}(f) are identical, while the experimental curves are different for $\mathbf{H}\in{ac^*}$ and $\mathbf{H}\in{bc^*}$. 
This is explained by our $C_3$-symmetrized model, which suppresses the \textit{in-plane} anisotropy, whereas the real \rucl is apparently monoclinic.
Therefore, the agreement between theory and experiment is excellent only in \cref{RuCl3-1}(f).
Nevertheless, we obtain overall semi-quantiatve agreement between theory and experiment.

\Cref{RuCl3-1}(d) shows the field-angular dependence of the magnetization within the $ab$ honeycomb plane for moderate $\mu_0 H $=2\,T.
In the angular dependence, components with clear 2-fold and 6-fold symmetries are identified.
Assuming a honeycomb lattice with $C_6$ symmetry, as present in the proposed $R\bar 3$ structure of \rucl, only an angular dependence with 6-fold symmetry is expected.
A spontaneous selection of single-domain zigzag magnetic order can break the 6-fold symmetry and give a component with 2-fold symmetry.
However, the same angular dependence of $M$ as shown in \cref{RuCl3-1}(d) is reproduced repeatedly for every measurement, even after heating the sample to room temperature, well above $T_{\mathrm N}$. 
Hence the preference of zig-zag domain selection with respect to the crystallographic axes is consistent, and probably related to the crystal structure, compatible with the suggested monoclinic C2/m space group. 
Accordingly, the zero-field magnetic Hamiltonian favors certain zigzag domains energetically out of the three possible domain directions. 
From the measured angular preference we infer, analogously as done in Ref.~\cite{lampenkelley2018fieldinduced}, that in our sample the dominant domain at low field is that with ordering wave vector $\mathbf Q \parallel b$ (as illustrated in \cref{RuCl3-1}(a)).
While this domain is expected to stay stable at finite fields $\mathbf H \parallel b$, a re-orientation to the other zigzag domains is expected at an intermediate field when $\mathbf H \parallel a$ \cite{winte18,lampenkelley2018fieldinduced}.

    \begin{figure}
 	    %FIG #2
    \includegraphics[width=8.5truecm]{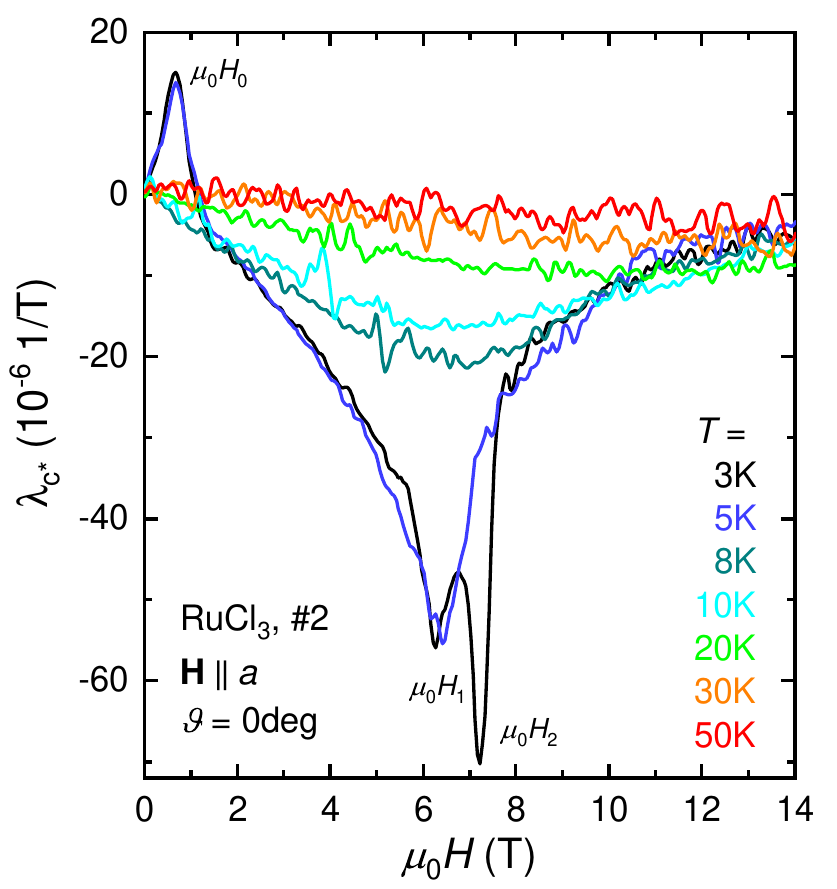}
    \caption{(Color online)
    Magnetic field dependence of the $\lambda_{c^*}$ linear magnetostriction coefficient at selected temperatures. The $\Delta{L}$ length change was measured along the $c^*$ axis and $\mathbf{H}\parallel{a}$ magnetic field was applied, perpendicular to one of the Ru-Ru bonds.}
    \label{RuCl3-2}
    \end{figure}

\begin{figure*}[t!]
    %FIG #3
\includegraphics[width=17truecm]{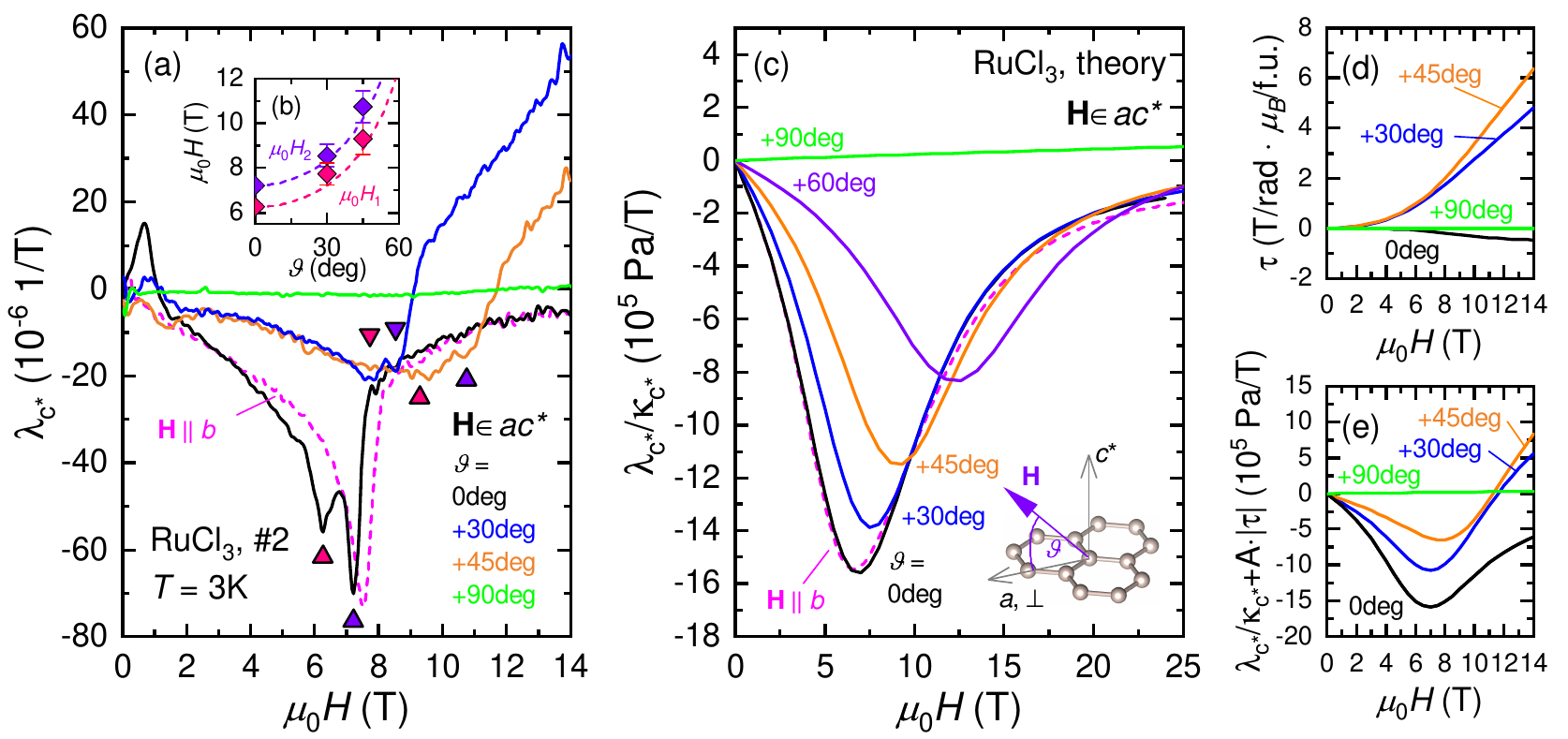}
\caption{(Color online)
(a) Field dependence of the linear magnetostriction coefficient $\lambda_{c^*}$ at $T$=3\,K. The magnetic field was applied along the main crystallographic axes, as well as canted out of the $ab$ plane with projection perpendicular to the bond ($\mathbf{H}\in ac^*$).
The $\vartheta$=0\,deg and 90\,deg angles correspond to the $a$ and $c^*$ axes, respectively.
The two peaks in the $\lambda_\cstar$ data correspond to ($\mu_0 H_1$) a phase transition between different zigzag interplane orderings and ($\mu_0 H_2$) a transition into the field-induced quantum paramagnetic phase. 
The $\mu_0H_1$ and $\mu_0H_2$ phase boundaries are indicated by triangles, respectively. 
(b) Angular dependence of the $\mu_0H_1$ and $\mu_0H_2$ phase transition fields. The dashed lines indicate the $\sim{1}/\cos{\vartheta}$ field dependence. 
(c) The field dependence of the $\lambda_{c^*}/\kappa_{c^*}$ magnetostriction calculated for $H$ field canted out of the $ab$ plane with $\vartheta$ angle, $\mathbf{H}\in ac^*$.
The inset defines the field angle $\vartheta$. Note, that panels (a) and (c) are shown for different field scales. 
(d) Magnetic field dependence of the calculated $\tau$ magnetic torque for selected $\vartheta$ canting angles, $\mathbf{H}\in ac^*$.
(e) The effect of magnetic torque on the field-dependence of the magnetostriction is modelled with the $\lambda_{c^*}/\kappa_{c^*} + A\cdot\vert\tau\vert$ relation with the same $A$=2.2$\cdot$10$^5$\,Pa$\cdot$Rad / (T$^2\cdot\mu_B$/f.u.) parameter fitted for each curve.
}
\label{RuCl3-3}
\end{figure*}

Figure~\ref{RuCl3-2} shows the experimental $c^*$-axis magnetostriction ($\lambda_{c^*}$, \cref{eq:magnetostrictionexp}) as a function of magnetic field $\mathbf H \parallel a$ %($\perp bond$) 
for selected temperatures within the ordered temperature regime, the short-range correlated Kitaev paramagnet, and the conventional thermal paramagnet. 
The magnetostrictions measured in increasing and decreasing fields were found to be identical within the accuracy of the measurement. 
At $T$=3\,K, the magnetostriction $\lambda_{c^*}$ has a positive peak at low fields and a sharp negative double-peak structure at higher fields.
The positive peak at $\mu_0H_0$=0.7\,T corresponds to the aforementioned domain re-population of the antiferromagnetic order and it is present in both the field increasing and decreasing runs.
We resolve two sharp negative peaks at $\mu_0H_1$=6.4\,T and $\mu_0H_2$=7.2\,T. 
The former is a phase transition at $\mu_0H_1$ where the inter-plane ordering between the zigzag-ordered honeycomb planes changes \cite{Balz2021intermediate}, while the latter at $\mu_0H_2$ is the transition where the zigzag magnetic order disappears.
In agreement with Refs.~\onlinecite{lampenkelley2018fieldinduced,bachus2021angle,Balz2021intermediate}, the extent between these two phases is the largest for $\mathbf H \parallel a$ and the smallest or absent for $\mathbf H \parallel b$ (cf.~\cref{RuCl3-3}(a)). 
At $T$=5\,K, the double-peak structure merges into a single, negative peak at $\mu_0H_1$=6.3\,T.
Above $T_{\rm N}=7.1$\,K, the sharp peaks of the low-temperature magnetostriction are replaced by broad field-dependent features.
For $T\gtrsim 30$\,K, the magnetostriction shows a linear field dependence, as expected for a conventional thermal paramagnet~\cite{Johannsen2005PRL}.
In contrast, for intermediate temperatures $T_{\rm N} < T \lesssim 30$\,K, we find the magnetostriction to show a non-linear and non-monotonic field dependence.
This appears to be a property of the short-range correlated Kitaev paramagnet~\cite{Do2017NatPhys,Jansa2018NatPhys,winte18,suzuki2020ProximateFerromagneticState} in this temperature range.

The $\lambda_{c^*}$ magnetostriction with fields applied along the main crystallographic axes as well as for $\mathbf H$ canted out from the $ab$ plane are shown in Fig.~\ref{RuCl3-3}(a).
The experimental configuration and the definition of the $\vartheta$ canting angle is illustrated in Fig.~\ref{RuCl3-3}(c);
The $H$ field is canted away from the $a$ axis by angle $\vartheta$ within the $a\cstar$-plane.
The measurement with $\vartheta$=0\,deg and 90\,deg corresponds to $\mathbf{H}\parallel{a}$ and $\mathbf{H}\parallel{c}^*$, respectively.
For better comparison and for the sake of completeness, we present the $\mathbf{H}\parallel b$ data reproduced after Ref.~\cite{Gass2020PRB} (dashed pink line in Fig.~\ref{RuCl3-3}(a)).
The $\lambda_{c^*}$ magnetostriction for $\mathbf{H}\parallel b$ has one single negative peak at $\mu_0H$=7.5\,T. 
Unlike $\lambda_\cstar$ for $\mathbf H\parallel a$, no significant domain re-orientation at low fields is visible, as expected for the identified dominant \mbox{$\mathbf Q\parallel b$} zigzag domain (Fig.~\ref{RuCl3-1}(a)). 
In contrast to the in-plane field results, the magnetostriction for $\mathbf{H}\parallel{c}^*$ is small and shows weak, non-monotonous field dependence.
The $\mu_0H_1$ and $\mu_0H_2$ critical fields of the two peaks in the magnetostriction data, approximately follow a simple $\sim{1}/\cos{\vartheta}$ angular dependence, as shown in Fig.~\ref{RuCl3-3}(b). 
Such an angular dependence is expected if the phase transitions are entirely driven by the in-plane component of the magnetic field.
We note, that the $\mu_0H_0$ critical field does not follow the $\sim{1}/\cos{\vartheta}$ angular dependence.
This deviation suggests that the reorientation of the three differently oriented zigzag domains in the presence of canted fields has a non-trivial energetic competition.

Theoretical calculations for the $\lambda_{c^*}/\kappa_{c^*}$ magnetostriction are shown in Fig.~\ref{RuCl3-3}(c) for the same field configurations as in Fig.~\ref{RuCl3-3}(a).
The calculated magnetostrictions for fields applied along the main crystallographic axes ($a,b,\cstar$) qualitatively reproduce the measured data.
In the experiments, we ascribed the measured $\mu_0H_0$=0.7\,T peak to the zig-zag domain reorientation. 
However, in our $C_3$-symmetrized model, there is no preferred domain orientation at $\mu_0H$=0\,T, and therefore no reorientation is expected.
Moreover, due to the restriction to a two-dimensional finite cluster in the calculations, the results are limited in the reproduction of the lower-field peak $\mu_0H_1$ for $\mathbf{H}\parallel a$ (related to the inter-layer re-ordering~\cite{Balz2021intermediate}), and peaks at phase transitions are generally expected to be broadened.
When the magnetic field is tilted out of the $ab$ plane, $\mathbf{H}\in ac^*$, theoretical calculations show that the peaks in the $\lambda_{c^*}/\kappa_{c^*}$ become smaller and appear at higher fields. 
While the experimental data in Fig.~\ref{RuCl3-3}(a) for $\vartheta$=30\,deg and 45\,deg retain the double-peak like features in $\lambda_{c^*}$, they differ significantly from the calculations.
In contrast to the theory, the measured $\lambda_{c^*}$ magnetostriction for $\theta$=30\,deg and 45\,deg changes sign at intermediate field strengths due to a large positive component added to the measurement.

We attribute the observed anomalous component to magnetic torque effects.
When the magnetic torque is strong, it could rotate, bend, and deform the \rucl crystal within the dilatometer, as discussed in the Supplementary, in Fig.~S2.
Theoretical calculations for the magnetic torque $\tau=\frac{dF}{d\vartheta}$ ($F$ being the free energy) for $\mathbf{H}\in ac^*$ are shown in Fig.~\ref{RuCl3-3}(d).
The torque, facilitated by $\Gamma$-exchange and $g$-anisotropy \cite{Riedl2019PRL}, is small for fields along the main crystallographic axes ($\vartheta=0$\,deg and $\vartheta=90$\,deg), but becomes large for intermediate canting angles where it strongly increases with field strength. 
Although the $H$ field points along a main crystallographic axis for $\vartheta$=0\,deg, ($\mathbf{H}\parallel a$), a small but nonzero torque persists anyway. 
Note that no symmetry in the Hamiltonian requires the torque to be maximal at $\theta$=45\,deg.
While the presently employed model parameters predict the magnetic torque to reach its maximum close to $\vartheta$=45\,deg in Fig.~\ref{RuCl3-3}(d), a smaller g-tensor anisotropy can further decrease the canting angle needed for maximum torque.
This can explain why the positive contribution in the  $\lambda_{c^*}$ magnetostriction measurement is larger for $\vartheta$=30\,deg than for $\vartheta$=45\,deg in Fig.~\ref{RuCl3-3}(a).
This further demonstrates that the effect of magnetic torque on the magnetostriction measurements is a complex issue, which depends on the spring constant, pressure setting, and dimensions of the dilatometer, as well as the dimensions and elastic constants of the sample.
For small rotations (deformations), it is reasonable to assume that the change in the magnetostriction is linear in the torque as $\Delta\lambda_{c^*}/\kappa_{c^*}\sim{A}\cdot\vert\tau\vert$, where $A$ is a material, measurement setup, and pressure setting dependent, but field magnitude- and angle-independent constant.
Figure~\ref{RuCl3-3}(e) illustrates the modified magnetostriction, calculated with the $\lambda_{c^*}/\kappa_{c^*}+{A}\cdot\vert\tau\vert$ relation, where $A$=2.2$\cdot$10$^5$\,Pa$\cdot$Rad / (T$^2\cdot\mu_B$/f.u.) is a fixed value for all curves.
The $A$ parameter was fitted to the $\vartheta$=45\,deg data with the highest magnetic torque $\tau$, so that $\lambda_{c^*}/\kappa_{c^*}(H^*) + A\cdot\vert\tau\vert=0$ is satisfied for the theoretical data at the same $H^*$ field as the measured magnetostriction ($\lambda_{c^*}(H^*)$=0).
This qualitatively models that the strong positive contributions to $\lambda_\cstar$ in the measurements with canted fields ($\vartheta$=+30\,deg and +45\,deg) are related to the rotational effect of the magnetic torque, and not intrinsic to the sample.
However, note that even with these efforts, the effect of torque cannot be removed from the measurement data in a quantitative manner.

\begin{figure*}
    \includegraphics[width=17truecm]{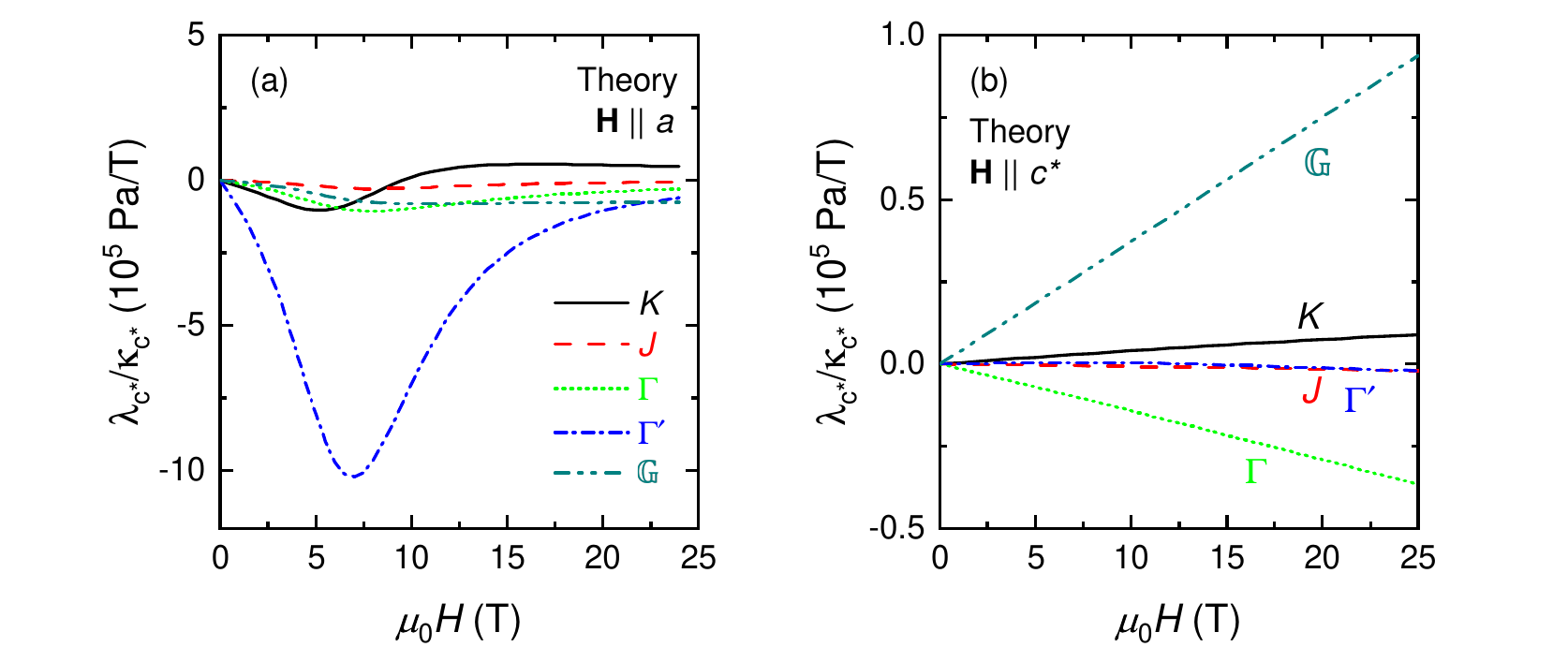}
    \caption{(Color online)
    Dissection of the largest contributions to theoretical magnetostriction through \cref{eq:lambda_theory}.
    Each curve correspond to a component of \cref{eq:lambda_theory} with different $\mathcal J$.
    The notation $K$, $J$, $\Gamma$, $\Gamma'$, and $\mathbb{G}$ corresponds to the summand components of $\frac{1}{V} \MEC K \left(\frac{\partial M}{\partial{K}}\right)$, $\frac{1}{V}\MEC J\left(\frac{\partial M}{\partial{J}}\right)$, $\frac{1}{V}\MEC \Gamma\left(\frac{\partial M}{\partial{\Gamma}}\right)$, $\frac{1}{V}\MEC \Gamma'\left(\frac{\partial M}{\partial{\Gamma'}}\right)$, and $\frac{1}{V}\widetilde{\mathbb{G}}\left(\frac{\partial M}{\partial{\mathbb{G}}}\right)$, respectively.
    The line with $\mathcal J = \mathbb G$ corresponds to magnetoelastic coupling with the $g$-tensor. (a)~In-plane field $\mathbf H \parallel a$ ($\vartheta=0$deg), (b)~Out-of-plane field $\mathbf H \parallel \cstar$ ($\vartheta=90$deg).
    }
    \label{fig:dissection} 
\end{figure*}

Focusing back on the crystallographic axes $a$ ($\vartheta$=0\,deg) and $\cstar$ ($\vartheta$=90\,deg), where torque effects are expected to be much weaker, we point out a much stronger anisotropy found in $\lambda_\cstar$ than expected from the magnetization anisotropy~\cite{Johannsen2005PRL}.
In principle, due to the Maxwell relation $\lambda_\cstar = - \partial M / \partial p_\cstar$, one can expect $\lambda_\cstar$ to be roughly proportional to the magnetization $M$ at small field strengths.
However, this does not explain the observed angular dependence of $\lambda_\cstar$:
While the magnetization for $\vartheta$=90\,deg is already reduced by a factor of $\sim 6$ to 10 compared to $\vartheta$=0\,deg (cf.~\cref{RuCl3-1}(e)), this alone cannot account for the much larger  $\sim 30$-fold reduction of the magnetostriction between $\vartheta$=90\,deg and $\vartheta$=0\,deg (cf.~\cref{RuCl3-3}(a)). 
This increased anisotropy effect is also reproduced in our model calculations (\cref{RuCl3-3}(c)).
In the calculations, we can trace the unusual reduction in magnetostriction back to contributions from different magnetoelastic couplings, i.e.~from different summands in \cref{eq:lambda_theory}. 
The largest entering magnetoelastic couplings $\MEC{\mathcal J}$ are the nearest-neighbor anisotropic couplings $\MEC K, \MEC \Gamma, \MEC \Gamma'$, which are field-independent. 
The field-strength and field-direction dependency enters through the susceptibilities $\partial M / \partial \mathcal J$. \Cref{fig:dissection}(a) shows the largest summands ($\MEC{\mathcal J} \cdot \partial M / \partial \mathcal J$) as a function of in-plane field $\mathbf H \parallel a$ ($\vartheta=0$deg), where a dominating effect from the contribution with $\mathcal J = \Gamma'$ is demonstrated. 
The large susceptibility $\partial M/\partial\Gamma'$ for in-plane fields can be understood from the fact that $\Gamma'$ is the exchange that tunes most strongly the easy-plane anisotropy of \rucl \cite{maksimov2020RethinkingRuCl}.
 However, for out-of-plane fields $\mathbf H \parallel \cstar$, variations in $\Gamma'$ have little effect onto the magnetization. Therefore the large $\MEC{\Gamma'}$-contribution breaks off for $\mathbf H \parallel \cstar$, leading to a much smaller $\lambda_\cstar$ as shown in  \cref{fig:dissection}(b). 
The agreement with experiment therefore confirms the presence of a strong \textit{negative} magnetoelastic $\MEC{\Gamma'}$ coupling in \rucl\ (see \cref{eq:MECmodel}).
Note that the large $\MEC\Gamma'<0$ suggests that the application of  3\% to 5\% compressive uniaxial \cstar-strain may destabilize the zigzag magnetic order~\cite{kaib2021magnetoelastic}.
While \rucl in measurements under hydrostatic pressure show dimerization~\cite{Biesner2018PRB,Bastien2018PRB}, the application of uniaxial strain leads to fundamentally different lattice deformations.  
As an example, compression along the \cstar axis expands the lattice within the ab plane, in contrast to the application of hydrostatic pressure, which compresses both the \cstar~axis and the honeycomb $ab$ plane. 
Another estimate for the uniaxial pressure dependence of $T_{\rm N}$ comes from the Ehrenfest relation~\cite{Johannsen2005PRL}:
\begin{equation}
\frac{\partial{T}_{\rm N}}{\partial{p}_\cstar}=V_{mol} T_{\rm N} \frac{\Delta{\alpha}_\cstar}{\Delta{C}_p},
\end{equation}
where $p_\cstar$ is the uniaxial pressure applied along the \cstar~axis, $V_{mol}$ is the molar volume, and $\Delta{C}_p$ and $\Delta{\alpha}_\cstar$ are the heights of the anomaly in the specific-heat and thermal-expansion at $T_{\rm N}$, respectively.
Using specific heat data $\Delta{C}_p$=3.0\,J/mol/K from Ref.~\onlinecite{Wolter2017PRB}, $\Delta{\alpha}_\cstar$=$-$7$\cdot$10$^{-5}$\,1/K from Ref.~\onlinecite{Gass2020PRB}, and $V_{mol}$=5.26$\cdot$10$^{-5}$\,m$^3$/mol, we get $\frac{\partial{T}_{\rm N}}{\partial{p}_\cstar}\approx-$8.8\,K/GPa.
This means that a noticeable change in $T_{\rm N}$ can be obtained under experimentally achievable conditions~\cite{Nakajima2015PRL}.

    \begin{figure*}[t!]
 	    %FIG #4
    \includegraphics[width=17truecm]{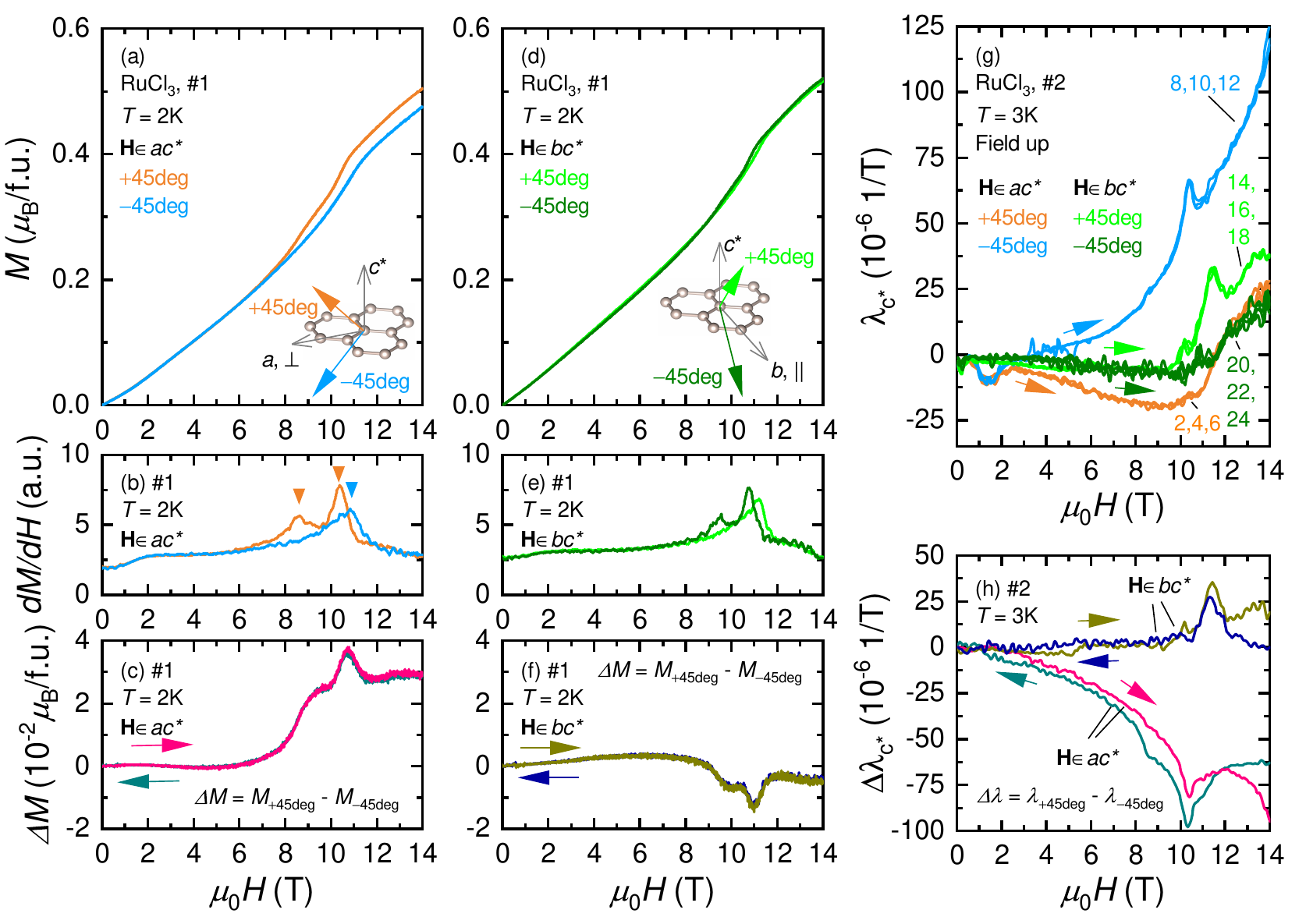}
    \caption{(Color online)
    (a) Magnetic field dependence of the magnetization at $T$=2\,K for $\mathbf{H}\in ac^*$ with $\vartheta$=$\pm$45\,deg canting out of the $ab$ plane.
    (b) Magnetic field dependence of the field-derivative, and (c) the field dependence of the $\Delta{M}$ magnetization difference for the $\vartheta$=$+$45\,deg and $-$45\,deg measurements ($\Delta{M}$=$M_\mathrm{+45\,deg}-M_\mathrm{-45\,deg}$).
    The phase transitions for $\mathbf{H}\in ac^*$ are indicated by triangles as peaks in $dM/dH$.
    (d-f) Magnetic field dependence of the magnetization, field derivative, and $\Delta{M}$ at $T$=2\,K for $\mathbf{H}\in bc^*$, $\vartheta$=$\pm$45\,deg. 
    (g) Magnetic field dependence of the $\lambda_{c^*}$ magnetostriction coefficient at $T$=3\,K for $\mathbf{H}\in ac^*$ and $\mathbf{H}\in bc^*$ measured in the field-increasing runs with $\vartheta$=$\pm$45\,deg canting angles. Three $\lambda_{c^*}$-$H$ curves are shown for each configuration, with numerals indicating the order of the measurements (complete list is shown in the supplementary material~\cite{Kocsis2021PRB2SM}, Fig.~\ref{RuCl3-1Suppl}).
	The experimental conditions are illustrated as an inset in panels (a) and (d).
    While panels (a,b,d,e,g) show measurements for the field-increasing runs, panels (c,f,h) show the measurements for the field increasing and decreasing runs. }
    \label{RuCl3-4}
    \end{figure*}

While the magnetostriction measurements under canted magnetic fields are strongly affected by the magnetic torque, the magnetization measurements are unaffected, as the sample is strongly fixed to a rigid sample holder.
Fig.~\ref{RuCl3-4} reveals an interesting anisotropy found in the magnetization measurements for fields rotated out of the $ab$-plane into opposite directions (i.e.~towards $+\cstar$ or $-\cstar$).
Magnetization measurements at $T$=2\,K for $\mathbf{H}\in ac^*$ and $\mathbf{H}\in bc^*$, canted out of the $ab$-plane by $\vartheta$=$\pm$45\,deg angles are shown in Figs.~\ref{RuCl3-4}(a-c) and \ref{RuCl3-4}(d-f), respectively.
Figures~\ref{RuCl3-4}(a,d), \ref{RuCl3-4}(b,e), and \ref{RuCl3-4}(c,f) show the $H$-field dependence of the magnetization, the field-derivative, and the field dependence of the $\Delta{M}$ magnetization difference between the $\vartheta$=$+$45\,deg and $-$45\,deg measurements ($\Delta{M}$=$M_\mathrm{+45\,deg}-M_\mathrm{-45\,deg}$), respectively.
For $\mathbf{H}\in ac^*$, in Fig.~\ref{RuCl3-4}(a), the field dependence of the magnetization for $\vartheta$=$+$45\,deg and $-$45\,deg show clear differences above $\mu_0H$=8\,T, while those of $\mathbf{H}\in bc^*$ show small differences only.
Moreover, peaks in the field derivative of the magnetization (Fig.~\ref{RuCl3-4}(b)) indicate two phase transitions for the $\vartheta$=$+$45\,deg measurement at $\mu_0H_1$=8.6\,T and $\mu_0H_2$=10.3\,T, while in the $\vartheta$=$-$45\,deg measurement there is only one peak seen at $\mu_0H'_2$=10.9\,T.
Note, that the field derivatives in the $\mathbf{H}\in bc^*$ measurements are similar to those of the $\mathbf{H}\in ac^*$ experiments, however here the $\vartheta$=$-$45\,deg  measurement has two peaks in $dM/dH$ and the $\vartheta$=$+$45\,deg measurement has one at slightly different fields.
The $\Delta{M}$ magnetization difference in Fig.~\ref{RuCl3-4}(c) shows a shoulder-like magnetization change starting from $\mu_0H$=8.5\,T and a peak at around $\mu_0H$=10.7\,T.
The $\Delta{M}$ magnetization difference for $\mathbf{H}\in bc^*$ in Fig.~\ref{RuCl3-4}(f) is about 4 times smaller and has opposite sign than those for $\mathbf{H}\in ac^*$.

For comparison, Fig.~\ref{RuCl3-4}(g) shows the field dependence of the $\lambda_{c^*}$ magnetostrictions for $\mathbf{H}\in ac^*$ and $\mathbf{H}\in bc^*$ fields canted out from the $ab$-plane in $\vartheta$=$\pm$45\,deg angles for the field-increasing runs.
The complete set of measurements for the field-increasing and decreasing runs are shown in Fig.~\ref{RuCl3-1Suppl}, while additional measurements on sample $\#$1 are shown in Fig.~\ref{RuCl3-3Suppl}~\cite{Kocsis2021PRB2SM}.
During the $\lambda_{c^*}$ - $H$ measurements the $H$ field was swept between $\pm$14\,T several times, then the sample was removed and rotated to the next measurement configuration.
Curves labeled as field up and field down refer to measurements in $H$ field with increasing or decreasing magnitudes, respectively.
In Figs.~\ref{RuCl3-4}(g), \ref{RuCl3-1Suppl}, \ref{RuCl3-2Suppl}, and \ref{RuCl3-3Suppl} we show three $\lambda_{c^*}$ - $H$ curves to demonstrate the signal to noise level, while Fig.~\ref{RuCl3-3Suppl} discusses the reproducibility when we change the $\vartheta$ canting angle.
Note, that these magnetostriction measurements in such canted fields show significant hysteresis. 
However, the difference between the +45\,deg and -45\,deg magnetostriction, $\Delta\lambda_{c^*}$=$\lambda_{c^*\mathrm{,+45\,deg}}-\lambda_{c^*\mathrm{,-45\,deg}}$ (Fig.~\ref{RuCl3-4}(h)), is found to be rather independent of the direction of the field sweep. 
Similarly to the magnetization measurements, the magnetostriction shows a significant difference $\Delta\lambda_\cstar$ for $\mathbf H \in a\cstar$, but a small one for $\mathbf H \in b\cstar$. 
Furthermore we can also observe a shoulder above $\mu_0H$=8\,T and a peak at $\mu_0H$=10.4\,T for $\mathbf{H}\in ac^*$. 
Note that the $\Delta{M}$ and $\Delta\lambda_{c^*}$ curves show a different field dependence at low fields, which is related to torque effects not compensated by the subtraction.

 \begin{figure*}
    %FIG #5
    \includegraphics[width=15.5truecm]{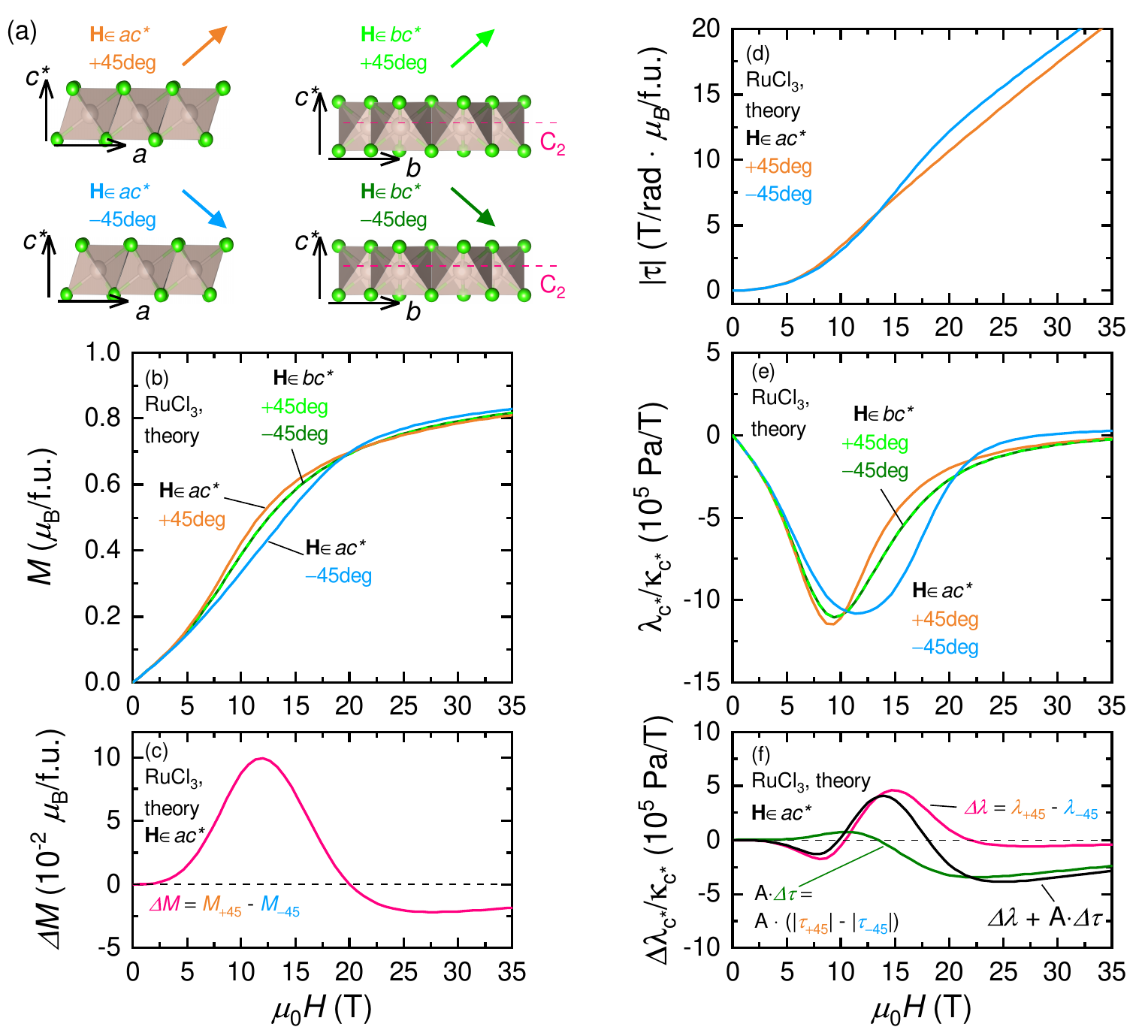}
    \caption{(Color online) 
    (a) Schematic illustration of the measurement configurations for each experiment. For $\mathbf{H}\in ac^*$, $\vartheta$=$+$45\,deg and $\vartheta$=$-$45\,deg, the net magnetization points along the vertexes and the edges of the Cl$_6$ octahedra, respectively.
    In both cases of $\mathbf{H}\in bc^*$, $\vartheta$=$\pm$45\,deg, the net magnetization points along the side of the RuCl$_6$ octahedra, which configurations are connected by the C$_2$ rotation in the honeycomb plane.
    (b) Magnetic field dependence of the magnetization and (c) $\Delta{M}$ magnetization difference calculated for $\mathbf{H}\in ac^*$ and $\mathbf{H}\in bc^*$ with $\vartheta$=$\pm$45\,deg canting out of the $ab$ plane.
    (d) Magnetic field dependence of the $\tau$ magnetic torque for $\mathbf{H}\in ac^*$, $\vartheta$=$\pm$45\,deg.
    (e) Magnetic field dependence of $\lambda_{c^*}/\kappa_{c^*}$ magnetostriction and (f) $\Delta\lambda_{c^*}/\kappa_{c^*}$ magnetostriction difference, calculated for $\mathbf{H}\in ac^*$ and $\mathbf{H}\in bc^*$, $\vartheta$=$\pm$45\,deg.
    The effect of the magnetic torque on the $\Delta\lambda_{c^*}/\kappa_{c^*}$ magnetostriction difference is illustrated in panel (f), the torque is scaled to the unit of $\lambda_{c^*}/\kappa_{c^*}$ with $A$=2.2$\cdot$10$^5$\,Pa$\cdot$Rad / (T$^2\cdot\mu_B$/f.u.).
    }
    \label{RuCl3-5}
    \end{figure*}

Theoretical calculations for the field dependence of the magnetization are shown in Fig.~\ref{RuCl3-5}(b) with the experimental configurations illustrated in Fig.~\ref{RuCl3-5}(a) for $\mathbf{H}\in ac^*$ as well as for $\mathbf{H}\in bc^*$ with fields canted out of the $ab$-plane by $\vartheta$=$\pm$45\,deg.
Note that the theoretical calculations are plotted on a wider field range than those of the measurements. 
In line with the experimental observations, the theoretical calculations confirm the different $M$-$H$ curves between the $\vartheta$=$+$45\,deg and $-$45\,deg canting angles for $\mathbf{H}\in ac^*$, and the calculated $\Delta{M}$ is shown in Fig.~\ref{RuCl3-5}(c). 
This non-symmetric difference in the magnetization is related to the orientation of the Ru$^{3+}$ magnetic moment within the Cl$_6$ octahedra, schematically illustrated in Fig.~\ref{RuCl3-5}(a).
For $\vartheta$=$+$45\,deg and $-$45\,deg angles with $\mathbf{H}\in ac^*$, the net magnetization points roughly towards the top vertex, or towards the midpoint of the edge of the RuCl$_6$ octahedra, respectively.
For $\mathbf{H}\in bc^*$, the calculated $M$-$H$ curves are exactly the same for the $\vartheta$=$+$45\,deg and $-$45\,deg cases.
In these cases, the Ru$^{3+}$ magnetic moments are pointing towards another edge of the RuCl$_6$ octahedra in sideways.
Both cases are connected by the C$_2$ rotation symmetry around the $b$ axis and therefore yield identical response in an ideal crystal.
The small $\Delta{M}$ difference observed for the $\mathbf{H}\in bc^*$ measurements can be related to twinning faults, where the honeycomb layers are rotated by 30\,deg with respect to each other.
This twinning fault is a different structural defect from the earlier recognized ABC/ABAB-stacking faults~\cite{Sears2015PRB,Banerjee2016NatMat}.

Calculations for the field dependence of the magnetic torque $\tau$, magnetostriction  $\lambda_{c^*}/\kappa_{c^*}$, and the magnetostriction difference $\Delta\lambda_{c^*}/\kappa_{c^*}$ for $\mathbf{H}\in ac^*$ and $\vartheta$=$\pm$45\,deg canting angles are shown in Fig.~\ref{RuCl3-5}(d), \ref{RuCl3-5}(e), and \ref{RuCl3-5}(f), respectively. 
Similarly to the experimental observations, the field dependence of $\lambda_{c^*}/\kappa_{c^*}$ is different for $\vartheta$=$+$45\,deg and $-$45\,deg, and $\Delta\lambda_{c^*}/\kappa_{c^*}$ is finite for $\mathbf{H}\in ac^*$. 
For $\mathbf{H}\in bc^*$ and $\vartheta$=$\pm$45\,deg, the $\lambda_{c^*}/\kappa_{c^*}$ curves are identical, similarly to the magnetization.
In order to account for the effect of the magnetic torque on the experimental data, Fig.~\ref{RuCl3-5}(d) shows calculations for the $\tau$ magnetic torque for $\mathbf{H}\in ac^*$, $\vartheta$=$\pm$45\,deg.
While the magnetic torque for both $\vartheta$=$\pm$45\,deg is large, the calculations show only slight differences in the magnitudes of the magnetic torques, \textit{i.e.} the $A\cdot\Delta\tau$=$A\cdot\vert\tau_{\rm +45\,deg}-\tau_{\rm -45\,deg}\vert$ is relatively small.
Still, we find that the modelled $A\cdot\Delta\tau$ is comparable in magnitude to $\Delta\lambda_{c^*}/\kappa_{c^*}$, which is shown in Fig.~\ref{RuCl3-5}(f) with the same scaling factor as in Fig.~\ref{RuCl3-3}(e) for $\vartheta$=$+$45\,deg .
This means that it is not possible to subtract the effect of torque on the measurements by simply measuring and subtracting the magnetostrictions in the $\pm\vartheta$ configurations.
Therefore, we consider the experimentally observed $\Delta\lambda_{c^*}$ in Fig.~\ref{RuCl3-4}(h) as an aggregate of the real $\vartheta$=$\pm$45\,deg non-symmetric anisotropies in the magnetostriction and a finite magnetic torque.
Additional theoretical calculations for the angular dependence of the magnetization for $\mathbf{H}\in ac^*$ and $\mathbf{H}\in bc^*$ are shown in Fig.~\ref{RuCl3-4Suppl}~\cite{Kocsis2021PRB2SM}.

%%%%%%%%%%%%%%%%%%%%%%%%%%%%%%%%%%%%%%%%%%%%%%%%%%%%%%%%%%%%%%%%%%%%%%%%%%%%%
%
%
%%%%%%%%%%%%%%%%%%%%%%%%%%%%%%%%%%%%%%%%%%%%%%%%%%%%%%%%%%%%%%%%%%%%%%%%%%%%%
	\section{Summary} \label{sec:Summary}
	We have studied the magnetic anisotropy in the Kitaev-candidate material \rucl, using field-dependent magnetization and magnetostriction $\lambda_{c^*}$ measurements.
	During these measurements, the magnetic field was selectively applied both along the main crystallographic axes or canted out from the honeycomb plane, while the length changes in the $\lambda_{c^*}$ experiments were always measured along the $c^*$ axis. 
	The field dependence of the low-temperature $\lambda_{c^*}$-$H$ magnetostriction measurements shows a double-peak structure for $\mathbf H \parallel a$ (perpendicular to one of the Ru-Ru bonds) and a single peak for $\mathbf H \parallel b$ (parallel to that Ru-Ru bond). 
	This is in agreement with the extents of the recently reported intermediate ordered phase with modified inter-plane ordering \cite{Balz2021intermediate,bachus2021angle}.

	We found that the $\lambda_{c^*}$-$H$ measurements show an unusually increased degree of field-angular anisotropy compared to the magnetization measurements ($\mathbf H \parallel a$ and $\mathbf H \parallel b$ experiments compared to $\mathbf H \parallel c^*$). 
	This suggests an additional degree of anisotropy in the magnetoelastic couplings.
	Our theoretical calculations based on \textit{ab-initio} derived magnetoelastic couplings show that this effect can be explained through the presence of a strong magnetoelastic $\MEC{\Gamma'}$-type coupling.
	The presence of the latter implies the possibility to destabilize the magnetic order via the application of uniaxial compressive strain.
	
	Both the $M$-$H$ and $\lambda_{c^*}$-$H$ measurements in the presence of canted fields show large differences and demonstrate a significant angular asymmetry when fields are canted away from the $a$-axis towards $+\cstar$ or $-\cstar$ axes ($\mathbf{H}\in{ac^*}$), i.e.~for positive or negative canting angles of the $H$ field.
	This angular asymmetry stems from the orientation of the $\mathbf H$ field with respect to the co-aligned RuCl$_6$ octahedra.
	However, we found that the magnetic torque has a strong influence on our magnetostriction measurements.
	From theory, magnetic torque is expected to be large only for canted field directions.
	We confirmed that the magnetic torque can qualitatively account for the measured field dependence of the magnetostriction and can contribute to the difference between of the magnetostrictions measured in positive and negative $\vartheta$ canting angles.
	This implies that when performing or comparing experiments in canted magnetic fields where the samples of different sizes are free-standing, such as in case of dilatometry or thermal Hall measurements, due to the very soft mechanical properties of \rucl the magnetic torque may add relevant contributions to the measurements via plastic distortion or tilting of the crystals.

\section*{Acknowledgements}
The authors are grateful for fruitful discussions with Taro Nakajima, Lukas Janssen, and Matthias Vojta.
The structural unit cell of the \rucl was illustrated using the software \texttt{VESTA}\cite{Momma2008}.
D.~G.~M. acknowledges support from the Gordon and Betty Moore Foundation’s EPiQS Initiative, Grant GBMF9069.
S.~N. was supported by the U.S. Department of Energy Office of Science, Division of Scientific User Facilities.
We acknowledge financial support from the German Research Foundation (DFG) through the Collaborative Research Center SFB 1143 (project-id 247310070), the W\"urzburg-Dresden Cluster of Excellence on Complexity and Topology in Quantum Matter ct.qmat (EXC 2147, project-id 390858490), and funding through DFG Project No. 411289067 (VA117/15-1) and TRR 288-422213477 (project A05).

%\bibliography{2021_pub_Kocsis_RuCl3-oblique_references.bib}


\newpage

\pagestyle{empty}

\newpage
\newpage

\renewcommand{\thefigure}{S\arabic{figure}}
\renewcommand{\theequation}{S\arabic{equation}}
\renewcommand{\thetable}{S\arabic{table}}
\setcounter{figure}{0}
\cleardoublepage

\begin{center}
\textbf{\boldmath Supplementary material for:\\ \lq\lq Investigation of the magnetoelastic coupling anisotropy in the Kitaev material $\alpha$-RuCl$_3$ \rq\rq \unboldmath}
\end{center}

    %***********************************************************************************************************************************************
    %FIG #1
    \begin{figure}[h!]
 	
    \includegraphics[width=8.5truecm]{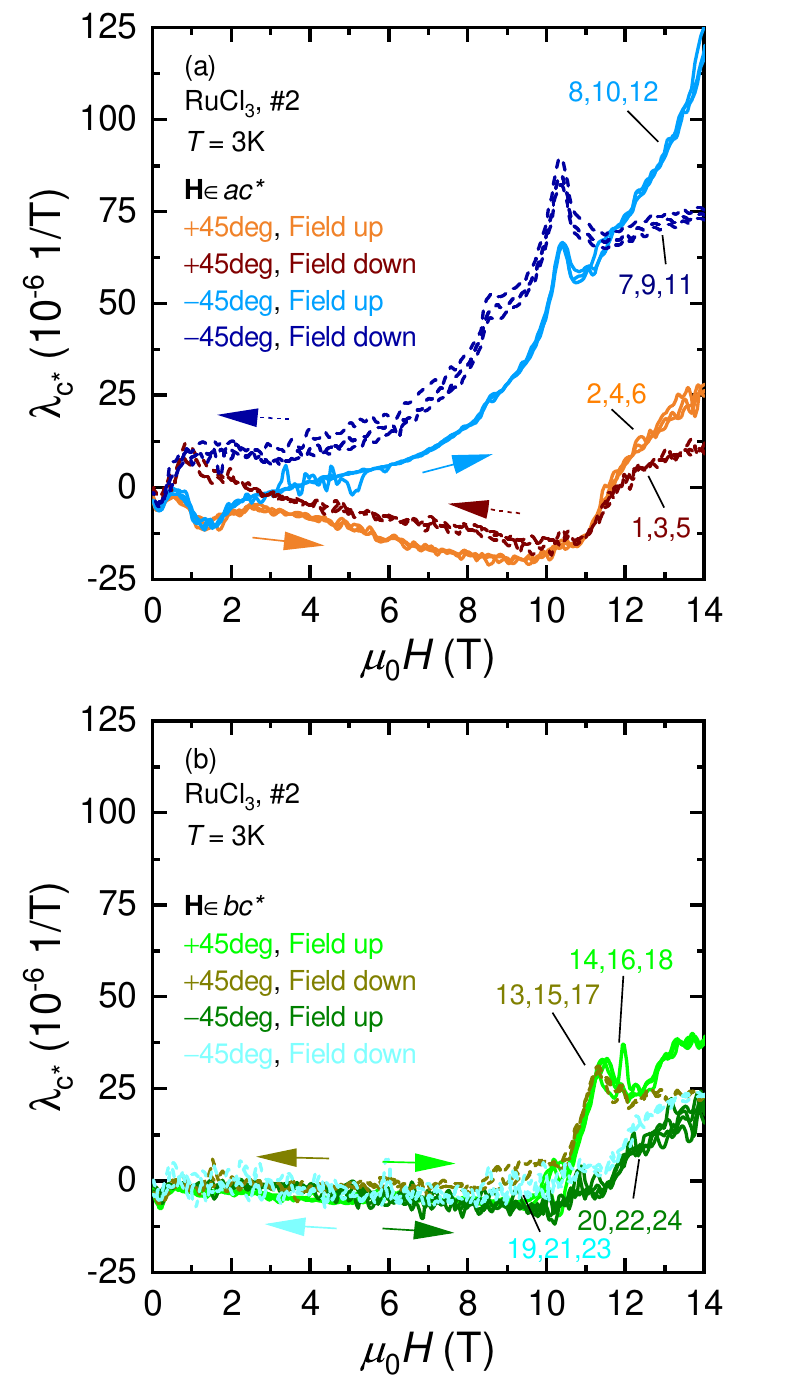}
    \caption{(Color online)
    Magnetic field dependence of the $\lambda_{c^*}$ linear magnetostriction coefficient at $T$=3\,K for the field-increasing and decreasing runs for field canted within the $ac*$ and $bc*$ planes ($\mathbf{H}\in{a}c*$ and $\mathbf{H}\in{b}c*$, respectively). For each setting, three measurements are shown, the numerals indicate the order of the measurements. The $ab$-plane projection of the canted magnetic field was set perpendicular and parallel to the Ru-Ru bonds, respectively in panels (a) and (b). The magnetic field was canted out of the $ab$ plane in $\vartheta$=$\pm$45\,deg angles.}
    \label{RuCl3-1Suppl}
    \end{figure}
    %***********************************************************************************************************************************************

Figure~\ref{RuCl3-1Suppl} shows the magnetic field dependence of the $\lambda_{c^*}$ linear magnetostriction coefficient at $T$=3\,K.
In panels (a) and (b), the $ab$-plane projection of the canted magnetic field was set parallel to the $a$ and $b$ axes, respectively.
The canting angle of the magnetic field was set to $\vartheta$=$\pm$45\,deg.
The $\lambda_{c^*}$-$H$ curves both for $\mathbf{H}\in{a}c*$ and for $\mathbf{H}\in{b}c*$ were measured in a complete $\pm$14\,T cycling of the $H$ field; meaning 0\,T $\rightarrow$~$+$14\,T $\rightarrow$~$-$14\,T $\rightarrow$~$ +$14\,T $\rightarrow\ldots$ .
In this sequence, the 0\,T $\rightarrow$~$+$14\,T runs often show transient features, and therefore these measurements are not presented.
The field increasing and decreasing runs for $\mathbf{H}\in{a}c*$ and $\mathbf{H}\in{b}c*$ are shown with solid and dashed lines, meaning measurements with increasing and decreasing magnitudes of the magnetic field, respectively.
The field increasing and decreasing runs show considerable, but reproducible hysteresis, which is probably related to the plastic deformation of RuCl$_3$.

\begin{figure}[h!]
 	
    \includegraphics[width=8.5truecm]{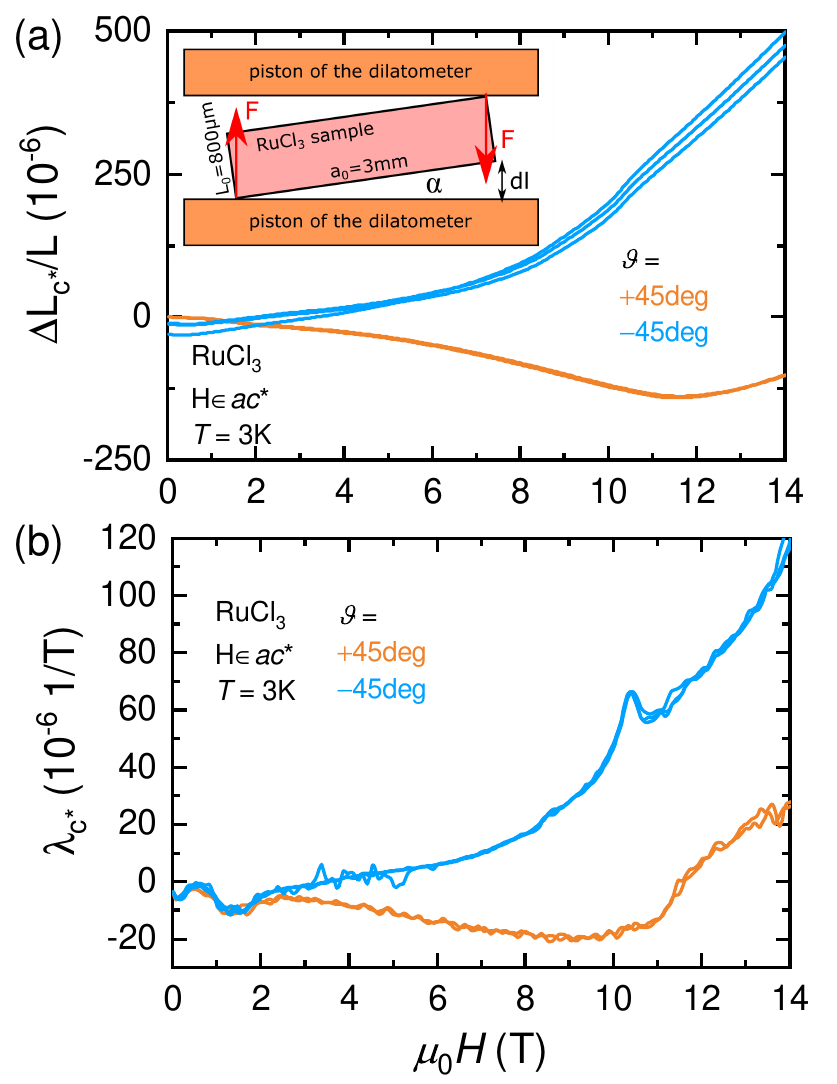}
    \caption{(Color online)
    Magnetic field dependence of the $\Delta{L}_{c*}/L$ relative length change and the corresponding $\lambda_{c^*}$ linear magnetostriction coefficient at $T$=3.0\,K for the field increasing runs. The magnetic field was canted out from the $ab$ plane in $\vartheta$=$\pm$45\,deg angles, perpendicular to the Ru-Ru bonds, $\mathbf{H}\in{ac^*}$. The inset of panel (a) illustrates the rotation of the sample within the dilatometer, where the rotation angle is $\alpha\approx$\,0.08\,deg.}
    \label{RuCl3-2Suppl}
    \end{figure}

Here we give an estimate for the effect of magnetic torque in the $\lambda_{c^*}$ magnetostriction measurements, using the directly measured $\Delta{L}_{c*}/L$ relative length change shown in Fig.~\ref{RuCl3-2Suppl}.
As discussed in the main text, in a magnetostriction measurement the sample is held in place in the dilatometer by a force $F$ applied on the sample during the mounting.
When the sample is placed in a magnetic field, which is canted away from the main crystallographic axes, a magnetic torque $\tau$ appears.
This magnetic torque tries to rotate the crystal within the dilatometer, as illustrated in the inset of Fig.~\ref{RuCl3-2Suppl}(a).
The sample rotates by a rotation angle $\alpha$ if the torque is larger than the $F\cdot{a_0}$ maximum torque exerted by the compression force of the dilatometer:
\begin{equation}
\tau\geq{F}\cdot{a_0},\label{eqS1}
\end{equation}
where the lateral size of our samples are $a_0$=3.0\,mm.
The rotation angle $\alpha$ of the sample can be estimated using the measurements in Fig.~\ref{RuCl3-2Suppl}.
Using the directly measured $\Delta{L}_{c^*}/L$ in Fig.~\ref{RuCl3-2Suppl}(a), we estimate the positive contribution of the magnetic torque to be of the order of $\left(\frac{\Delta{L}_{c^*}}{L}\right)_\tau\sim$5$\cdot$10$^{-3}$, at the highest field, where the thickness of the sample is $L_0$=800\,$\mu$m.
The $\Delta{L}_{c^*}/L$ gives the estimation for the maximum of the $dl$ lifting of the side of the sample, as $dl\sim\left(\frac{\Delta{L}_{c^*}}{L}\right)_\tau L_0$.
Therefore, we can estimate the maximum of the $\alpha$ angle, as $sin(\alpha)=\frac{dl}{a_0}$, which is $\alpha_{\rm max}\approx\,$0.08\,deg.
Note, that according to Eq.~\ref{eqS1}, the effect of the magnetic torque can be reduced if the lateral size of the sample ($a_0$) or the force ($F$) are increased.

    %***********************************************************************************************************************************************
    %FIG #2
    %***********************************************************************************************************************************************

\begin{figure*}
 	
    \includegraphics[width=16truecm]{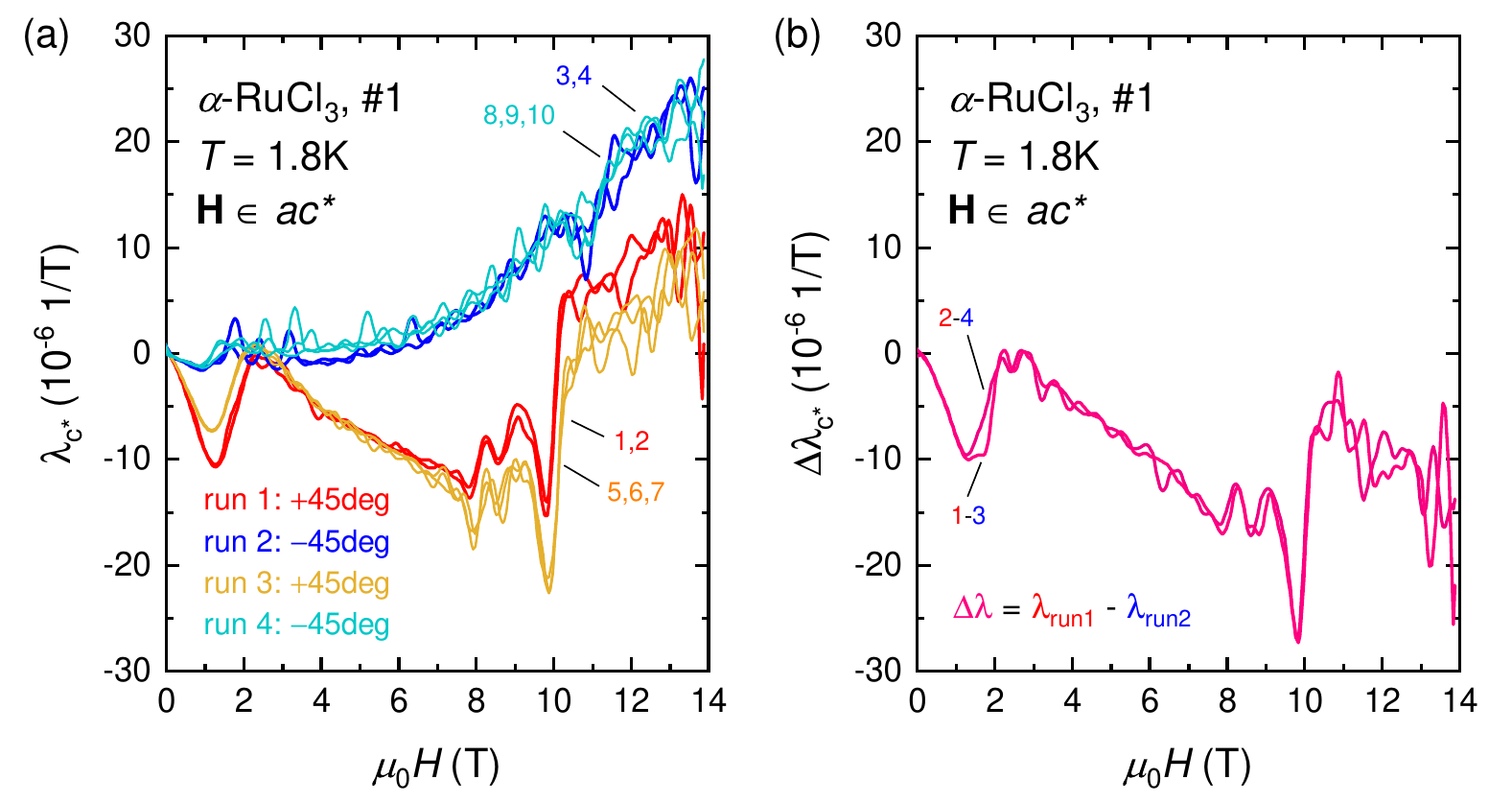}
    \caption{(Color online)
    (a) Magnetic field dependence of the $\lambda_{c^*}$ linear magnetostriction coefficient at $T$=1.8\,K for sample $\#$1. The magnetic field was canted out of the $ab$ plane in $\vartheta$=$\pm$45\,deg angles, $\mathbf{H}\in{ac^*}$. During the measurement we changed the $\vartheta$ canting angle in a $+$45\,deg $\rightarrow$ $-$45\,deg $\rightarrow$ $+$45\,deg $\rightarrow$ $-$45\,deg sequence. Numerals indicate the order of the measurements.
    (b) Magnetic field dependence of the $\Delta\lambda_{c^*}$ magnetostriction difference calculated from the measurements runs 1 and 2.}
    \label{RuCl3-3Suppl}
    \end{figure*}

Figure~\ref{RuCl3-3Suppl} investigates the reproducibility of the $\lambda_{c^*}$ magnetostriction measurements for $\mathbf{H}\in{ac^*}$ and $\vartheta$=$\pm$45\,deg, using Sample $\#$1 at $T$=1.8\,K.
In this experiment, we have used a new dilatometer (\textit{Mini Dilatometer}, \textit{Kuechler}), which is different from the custom built one, used in the main text.
In this dilatometer it is possible to rotate the sample in-situ, which means that it is not necessary to remove the sample from the dilatometer to change the angle $\vartheta$.
Therefore, the force applied on the sample is preserved for each consecutive experiment.
To suppress the magnetic torque related signal, we have applied much larger force on Sample $\#$1 (in Fig.~\ref{RuCl3-3Suppl}), than that was applied in case of Sample $\#$2 (in Fig.~\ref{RuCl3-4}).
In this experiment, we measured the field dependence of the magnetostriction, $\lambda_{c^*}$-$H$ and changed the $\vartheta$ canting angle ($\mathbf{H}\in{ac^*}$) between $\pm$45\,deg four times.
We have found that the difference in $\lambda_{c^*}$-$H$ between the $\vartheta$ = $+$45\,deg and $-$45\,deg measurements qualitatively well reproduces in this experiment; and it shows a similar difference between the $\pm$45\,deg measurements as in Fig.~\ref{RuCl3-4}, however with a smaller magnitude.
Note also, that the magnetic torque related component is significantly reduced due to the higher compression force in the dilatometer.
However, the new measurements show also differences compared to the earlier measurements, \textit{i.e.} the magnitude of $\lambda_{c^*}$-$H$ is half as large than earlier and the phase transition peaks are sharper.
Moreover, the negative peak at $\mu_0{H}_{0}$=1.5\,T only appears for the $\vartheta$ = $+$45\,deg measurements and not for $\vartheta$ = $-$45\,deg, which is a different behavior compared to that shown in Fig.~\ref{RuCl3-4}(g,h).
These changes we relate to the larger uniaxial pressure applied during the measurement shown in Fig.~\ref{RuCl3-3Suppl} and to the lower measurement temperature ($T$ = 1.8\,K), respectively.
Figure~\ref{RuCl3-3Suppl}(b) shows the $\Delta\lambda_{c^*}$ magnetostriction difference calculated from the first two measurement runs.
The $\Delta\lambda_{c^*}$ shows a similar decrease in magnitude as the $\lambda_{c^*}$-$H$.
These differences motivate further, uniaxial compression dependent magnetization and magnetostriction measurements.

\begin{figure*}
 	
    \includegraphics[width=16truecm]{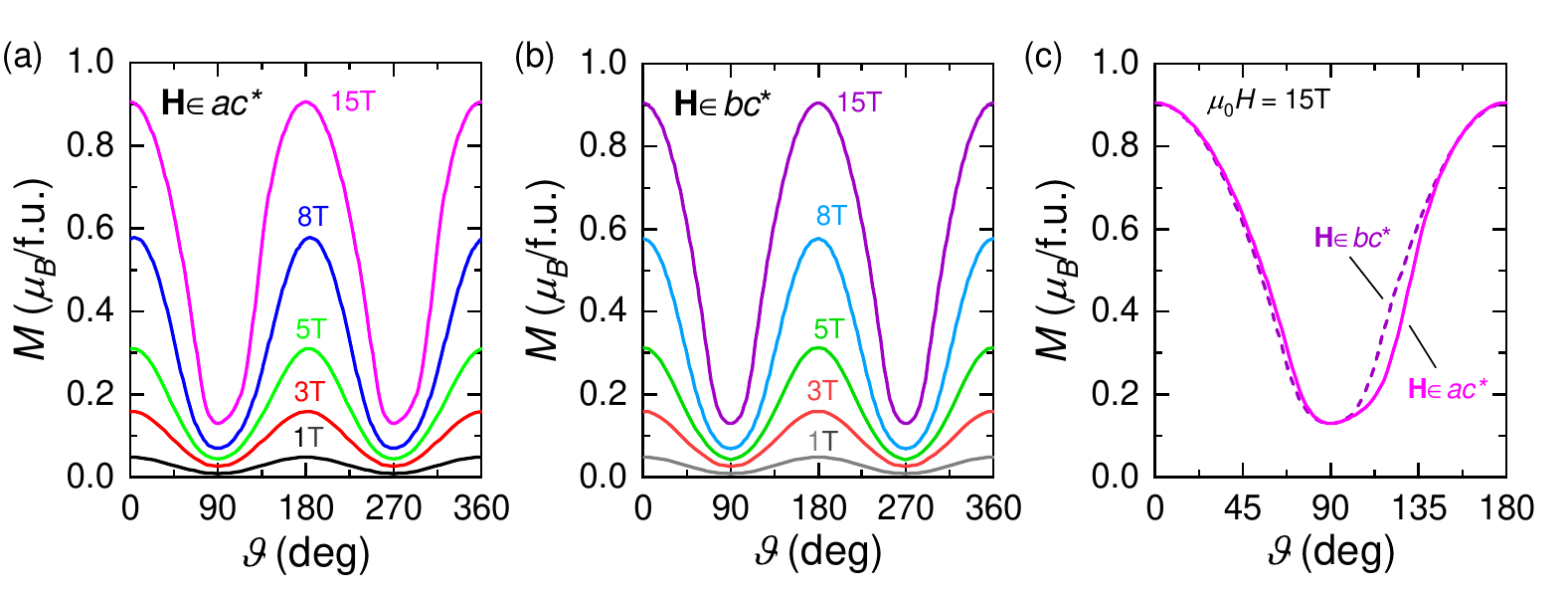}
    \caption{(Color online)
    Theoretical calculations for the angular dependence of the magnetization for (a) $\mathbf{H}\in{ac^*}$ and for (b) $\mathbf{H}\in{bc^*}$. In panel (c), we compare the angular dependency for $\mathbf{H}\in{ac^*}$ and $\mathbf{H}\in{bc^*}$ at $\mu_0H$=15\,T.}
    \label{RuCl3-4Suppl}
    \end{figure*}

In Fig.~\ref{RuCl3-4Suppl}(a,b) we show additional theoretical calculations for the angular dependence of the magnetization up to fields $\mu_0H$ = 15\,T for $\mathbf{H}\in{ac^*}$ and $\mathbf{H}\in{bc^*}$, respectively.
In Fig.~\ref{RuCl3-4Suppl}(c) we compare the angular dependence of the magnetization for $\mathbf{H}\in{ac^*}$ and $\mathbf{H}\in{bc^*}$ in the presence of $\mu_0H$ = 15\,T, the horizontal axis is expanded for better visibility.
Note that while the magnetization curve for $\mathbf{H}\in{bc^*}$ is symmetric with respect to the $c^*$ axis ($\vartheta$ = 90\,deg), the $M$-$\vartheta$ curve for $\mathbf{H}\in{ac^*}$ is clearly not symmetric.

    %***********************************************************************************************************************************************
    %FIG #3
    %***********************************************************************************************************************************************

    %***********************************************************************************************************************************************
    %FIG #4
    %***********************************************************************************************************************************************

\end{document}